\documentclass[structabstract]{aa}  
\usepackage[authoryear]{natbib}
\usepackage{graphicx}
\usepackage{amsmath}
\usepackage{amssymb}
\usepackage{txfonts}
\begin{document}
   \title{Continuous monitoring of pulse period variations in Her X-1
   using \textsl{Swift/BAT}}


   \author{D. Klochkov\inst{1}\and
           R. Staubert\inst{1}\and
           K. Postnov\inst{2}\and
           N. Shakura\inst{2}\and
           A. Santangelo\inst{1}
          }

   \institute{Institut f\"ur Astronomie und Astrophysik, Universit\"at 
     T\"ubingen (IAAT), Sand 1, 72076 T\"ubingen, Germany
     \and
     Sternberg Astronomical Institute, Moscow University, 119992 Moscow, Russia
   }

   \date{Received ***; accepted ***}

 
  \abstract
   {Monitoring of pulse period variations in accreting binary pulsars is an 
     important tool to study the interaction between the magnetosphere of
     the neutron star and the accretion disk. While the X-ray flux
     of the brightest X-ray pulsars have been successfully monitored over many
     years (e.g. with \textsl{RXTE/ASM, CGRO/BATSE, Swift/BAT}), 
     the possibility to monitor their pulse timing 
     properties continuously has so far been 
     very limited.}
   {In our work we show that the \textsl{Swift/BAT} observations can 
     be used to monitor coherent pulsations of bright X-ray sources 
     and use the \textsl{Swift} archival data to study one of the most 
     enigmatic X-ray pulsars, Hercules X-1.
    A quasi-continuous monitoring of the pulse
    period and the pulse period derivative of an X-ray pulsar, 
    here \hbox{Her~X-1}, is achieved over a long time ($\gtrsim 4$\,yrs).
     We compare our observational results with predictions of accretion 
     theory and use them to test different aspects of the physical model 
     of the system.}
   {In our analysis we use the data accumulated with \textsl{Swift/BAT} 
    starting 
    from the beginning of 2005 (shortly after launch) until the present time. 
    To search for pulsations and for their subsequent analysis we used the 
    count rate measured by the \textsl{BAT} 
    detector in the entire field of view.}
   {The slope of the correlation between the locally determined spin-up rate
   and the X-ray luminosity is measured for \hbox{Her X-1} and found to be 
   in agreement with predictions of basic accretion torque theory. 
   The observed behaviour of the pulse period together with the previously
   measured secular decrease of the system's orbital period is discussed
   in the frame of a model assuming ejection of matter close to the
   inner boundary of the accretion disk.}
   {}


   \maketitle
%

\section{Introduction}

The persistent accreting pulsar Hercules X-1 was one of the first
X-ray sources discovered by the \textsl{Uhuru} satellite in 1972
\citep{Tananbaum_etal72,Giacconi_etal73} and since then it remains
one of the most intensively studied X-ray pulsars. The basic
phenomenological picture of Her X-1 was established soon after its discovery:
a close binary system consisting of an accreting magnetized neutron star with 
a 1.24\,s spin period and a stellar companion HZ Her \citep[first suggested 
by][]{Liller72} -- a main sequence star of the spectral type A/F 
\citep{Crampton74}. The mass of the optical companion is $\sim$2$M_{\odot}$ 
which places the system in the middle between high and low mass X-ray 
binaries. Other main parameters of the binary system are the following:
orbital period $P_\text{orb}\simeq 1.7$\,days, 
X-ray luminosity of the 
source $L_{\rm X}\sim 2\times 10^{37} {\rm erg\,s}^{-1}$ for a distance of 
$\sim$7~kpc \citep{Reynolds_etal97}. The binary orbit is almost 
circular \citep{Staubert_etal09}
and has an inclination $i\sim 85-88^{\circ}$ \citep{GerendBoynton76}.
The magnetic field strength on the surface of the neutron star is 
believed to be
around $3\times 10^{12}$\,G, as estimated from the energy of the cyclotron
resonant scattering feature \citep{Truemper_etal78}.

Like many other X-ray pulsars Her X-1 exhibits significant 
variation of the pulsation period (i.e. spin rate of the neutron star).
Alternation of spin-up and spin-down episodes on time-scales of several
months in this system is superimposed on a background of systematic spin-up 
\citep{Sheffer_etal92,Staubert_etal06,Klochkov07}. The behaviour of the
pulsar's spin period on shorter time scales is not very well studied because
such a study would require a continuous monitoring of \hbox{Her X-1} with a
sensitive X-ray detector. Only between 1991 and 2000 the \textsl{BATSE} 
instrument onboard \textsl{CGRO} \citep{Fishman_etal89} 
provided the information about the source's pulse period on a regular basis. 
These data allowed \citet{Staubert_etal06} to reveal an anticorrelation
between the pulse period and times of X-ray \textsl{turn-ons}, 
i. e. switching from an \textsl{off}-state with low X-ray flux to the 
so-called \textsl{main-on}
state with high flux (such turn-ons regularly occur in \hbox{Her X-1} 
with a period of $\sim$35\,days and are
believed to be caused by a precessing tilted accretion disk around the
neutron star, see e.g. \citealt{GerendBoynton76}). 

In this work we present a continuous monitoring of the \hbox{Her X-1} pulse
period $P$ and its local (measured at the time of the observation) 
time derivative $\dot P$ using the \textsl{Swift/BAT} instrument
starting from 2005 (begin of scientific operation) to 2009. 
We compare the observed pulse period development with that measured
previously with \textsl{CGRO/BATSE}.
The data of the \hbox{monitoring} allowed us to explore for the first time the
correlation between the {\em locally} measured $\dot P$ and the X-ray flux
of the pulsar and compare the results with predictions of the accretion
theory. The observed strong spin-down episodes are discussed in the frame 
of a model assuming ejection of matter from the inner part of the 
accretion disk along the open magnetic field lines.


\section{Observations}

For our analysis we used the public archival data obtained with the 
\textbf{B}urst \textbf{A}lert \textbf{T}elescope 
(\textsl{BAT}, 150--150\,keV, \citealt{Barthelmy_etal06}) 
onboard the \textsl{Swift} observatory \citep{Gehrels_etal04}.
With its large field of view (1.4 sterad) the \textsl{BAT} instrument 
is originally designed to provide fast triggers for gamma-ray bursts 
and their accurate positions in the sky ($\sim$4\,arcsec). 
Following such a trigger, the observatory points in the direction of the burst,
which can be then observed with the X-ray and UV/optical telescopes
onboard the satellite. While searching for bursts, \textsl{BAT} 
points at different locations of the sky, thus, performing 
an all-sky monitoring in hard X-rays (measurements of the X-ray flux
are provided by the \textsl{Swift/BAT} team in the form of X-ray light curves
for the several hundred bright persistent and transient 
sources\footnote{http://heasarc.gsfc.nasa.gov/docs/swift/results/transients/}).

Most of the \textsl{BAT} observations are stored in the form of 
detector plane maps (histograms) accumulated over the 5-minutes exposure 
time which limits the possibilities of timing 
analysis. In addition to the detector plane histograms, however, the
stored data contain 64-msec count rates corresponding
to the total flux detected by \textsl{BAT}. If a bright pulsating source
with a known period falls into the field of view of the instrument, 
the total count rate can be used to search for coherent pulsations of 
that source. We have used this strategy to measure the 1.24\,s pulsations
of \hbox{Her X-1} during its main-on states (when the X-ray flux of the 
source is high) repeating every $\sim$35 days.


\section{Timing analysis of \textsl{BAT} data}

To determine the pulse period, we used the method similar to that
described by \citet{Staubert_etal09}. The method includes two techniques 
for the determination of the period: \textsl{epoch folding} 
with $\chi^2$ search 
\citep[e.g.][]{Leahy_etal83} and \textsl{pulse phase connection}
\citep[e.g.][]{Deeter_etal81}. The first one is used to establish the
presence of the periodic signal from \hbox{Her X-1} in the \textsl{BAT} data,
determine the approximate period, and construct pulse profiles 
(by folding the data with the found period), while the second 
is subsequently applied to the pulse profiles to determine the
precise value of the period and its time derivative.

\begin{figure}
  \resizebox{\hsize}{!}{\includegraphics[angle=-90]{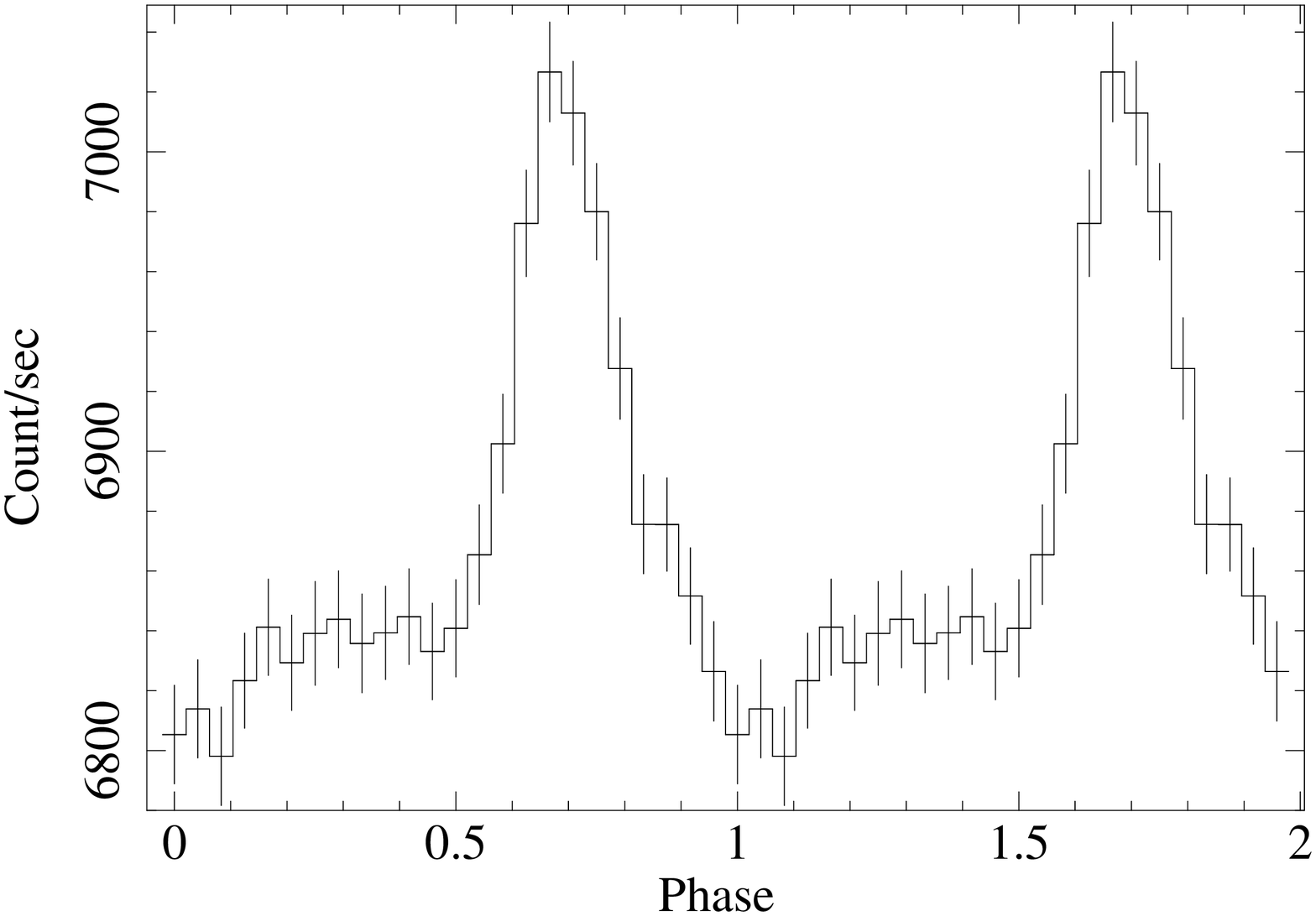}}
  \caption{An example of a \textsl{Swift/BAT} pulse profile used to study
    pulse shifts for the precise measurement of the pulse period variations
    in \hbox{Her X-1}. The profile is taken at MJD $54516.083$ with 
    $\sim$1200\,s integration time.
  }
  \label{profile}
\end{figure}

As already mentioned, for our analysis we used the total count rates 
measured by \textsl{BAT} with a time resolution of 64 msec. All times 
of the count rates were translated to the solar system barycenter and 
corrected for binary motion (using our newest orbital ephemeris presented in
\citealt{Staubert_etal09}). Then we performed a period search using
\textsl{epoch folding} in a narrow period interval around the expected
pulse period ($\sim$1.237\,s). If a strong periodic signal is present
we determined the period and used it to construct X-ray pulse profiles for 
subsequent \textsl{pulse phase connection}. The integration time in each
case was chosen to be large enough to obtain a pulse profile with sufficient
statistics (normally $\sim$1\,ksec). A typical profile used in our 
phase-connection analysis is shown in Fig.\,\ref{profile}. 

To obtain pulse arrival times for the subsequent phase connection analysis
we have fit all individual pulse profiles with a template profile 
constructed by superposing the individual profiles from the same main-on
state. In many cases we ignored the data at the start and the end
of a main-on where the profiles deviate noticeably from those obtained
in the middle of the main-on. Thus, within one main-on we restricted
our analysis to the time intervals where the shape of the pulse profile
does not change appreciably. Nevertheless, we cannot completely exclude
possible systematics which might affect our results 
due the variation of the profile 
shape (see also Discussion). If the interval between the adjacent
pulse arrival times is short enough (or if the uncertainty of the 
assumed pulse period is sufficiently small), one can reconstruct the number
of pulsation cycles passed in between excluding any mis-counting. In this
case, the estimated pulse arrival times can be analytically modelled. 
For example, if the pulse period $P$ is constant, the expected arrival 
time of pulse number $n$ is 
\begin{equation}
t_n = t_0+nP,
\label{tn_cons}
\end{equation}
where $t_0$ is the arrival time of the "zero"-th pulse. 
In case of non-zero 
first and second derivatives of the pulse period, the arrival times 
are given by \citep[see e.g.][]{Staubert_etal09}
\begin{equation}
t_n = t_0+nP_0+\frac{1}{2}n^2P\dot P_0+\frac{1}{6}n^2P_0^2\ddot P_0 + ...,
\label{tn_pdd}
\end{equation}
where $P_0$, $\dot P_0$, and $\ddot P_0$ are the pulse period and its 
time derivatives at the time $t_0$ respectively. 

\begin{figure}
  \resizebox{\hsize}{!}{\includegraphics{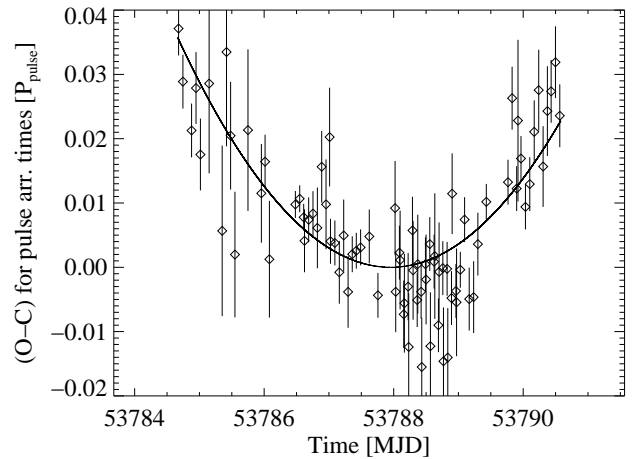}}
  \caption{Estimated (observed) minus calculated (assuming a constant period) 
    pulse arrival times of \hbox{Her X-1} in units of its pulse period 
    during one of its main-on states observed with \textsl{Swift/BAT}.
    Parabolic fit to the data shown with the solid line corresponds to
    a constant positive $\dot P\simeq 1.4\times 10^{-12}$\,s/s 
    (according to Eq.\,\ref{tn_pdd}).
  }
  \label{oc}
\end{figure}

\begin{figure*}
  \sidecaption
  \includegraphics[width=12cm]{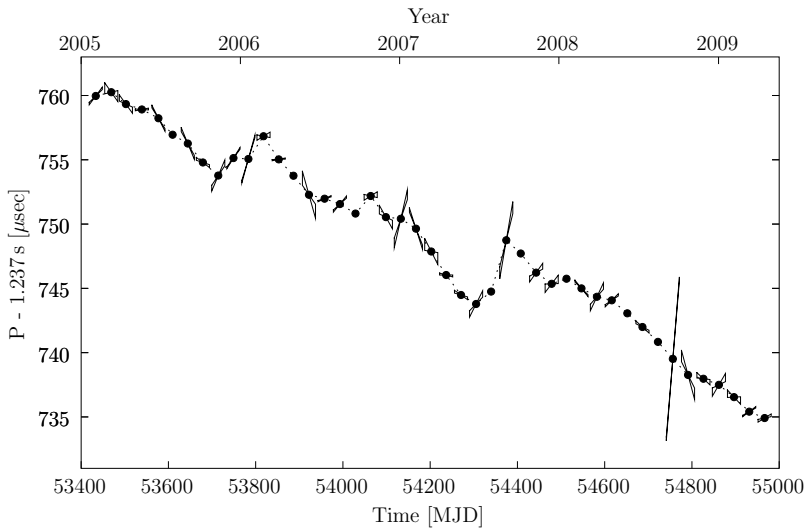}
  \caption{Pulse period $P$ of Her X-1 measured with \textsl{Swift/BAT} as 
  a function of time. The cones around each point indicate the allowed range 
  of the slope corresponding
  to the measured $\dot P$ and its uncertainties. Measurement errors
  of the period itself are smaller than the symbol sizes.}
  \label{per}
\end{figure*}

A convenient way to explore the variation of the pulse period,
often used in phase connection technique, 
is to construct the so-called $(O-C)$ diagram
showing the estimated (observed) pulse arrival time minus the
calculated one assuming a constant period (i.e. using 
Eq.\,\ref{tn_cons}). An example of such a diagram
measured with \textsl{BAT} during one of the \hbox{Her X-1} 
main-on states is shown in Fig.\,\ref{oc}. A straight line in the
graph would correspond to a constant period defined by the slope
of the line. The solid curve indicates
the parabolic fit to the data corresponding to a constant
positive $\dot P$ according to Eq. \,(\ref{tn_pdd}) (in the shown
case the best-fit $\dot P \simeq 1.4\times 10^{-12}$\,s/s).


\section{Pulse period variability}

Using the method described in the previous section we determined the
pulse period $P$ and its time derivative $\dot P$ for most of the 
\hbox{Her X-1} main-on cycles observed with \textsl{BAT}
(for other parts of the 35\,d cycle
the flux was too low for such determinations).
At the time of writing this paper the data are available for the time period
from March 2005 to May 2009 that covers 45 
35d cycles (main-on states) of the pulsar. 
For several cycles the \textsl{BAT} observations 
have relatively poor statistics due to gaps in the data. For such cycles
only $P$, but no  $\dot P$ could be determined.
For the cycle with the turn-on at MJD $\sim$54339.68
even the pulse period could not be found due to a gap in the data.
However, during this cycle \hbox{Her X-1} was observed with the
\textsl{INTEGRAL} satellite \citep{Staubert_etal09}. So, we included 
the value of $P$ measured with \textsl{INTEGRAL} in our data set 
(since the \textsl{INTEGRAL} observations caught only the end of the 
main-on, $\dot P$ could not be measured).  

For the main-on starting at MJD $\sim$53577.14 only the pulse period $P$
could be measured with \textsl{BAT} (no $\dot P$). But the main-on
was simultaneously observed with the \textsl{INTEGRAL} and \textsl{RXTE}
observatories \citep{Klochkov_etal08, Staubert_etal09}. The value
of $P$ measured during these pointings is consistent (within
1$\sigma$ uncertainty) with the value found from the \textsl{BAT} data.
We consider this agreement as a successful cross-check between the
\textsl{Swift}, \textsl{RXTE}, and \textsl{INTEGRAL} observations.

The time evolution of the measured pulse period for the main-ons of
\hbox{Her X-1} is shown in Fig.\,\ref{per}. Where a corresponding
value of $\dot P$ was measured, the 1$\sigma$-uncertainty range is indicated
by the cones, the orientation 
of which reproduce the measured $\dot P$ value.

A close quantitative inspection of all measured values of $P$ and 
$\dot P$ leads to the following list of statements about their 
evolution with time.
\begin{itemize}
\item The pulse period evolution shown in Fig.\,\ref{per} resembles
a saw-tooth where one can distinguish five spin-down and five spin-up 
episodes (from comparing adjacent measurements). 
The overall mean spin-up rate is 
$<\dot P> = -1.8203(3)\times10^{-13}$s\,s$^{-1}$
or $-15.728(2)$\,ns\,d$^{-1}$. We note that this is 
slightly steeper than the general mean 
spin-up of $-9$\,ns\,d$^{-1}$ observed from the discovery of the source to 
the dramatic spin-down event during the \textsl{Anomalous Low} of 
1999/2000 \citep{Klochkov07}.
\item The spin-down episodes (with mean spin-down rates between 
$0.95\times 10^{-13}$ and $4.5\times 10^{-13}$\,s\,s$^{-1}$), 
lasting from 1 to 3 35\,d cycles, are generally shorter than the spin-up 
episodes (with mean spin-up rates from $2.9\times 10^{-13}$  
to $4.4\times 10^{-13}$\,s\,s$^{-1}$) which last from 3 to 13
35\,d cycles.
\item Locally measured $\dot P$ values, ranging from $-11\times 10^{-13}$\,
s\,s$^{-1}$ to $+$$48\times 10^{-13}$\,s\,s$^{-1}$, show stronger modulation 
than $\dot P$ values found from comparing the pulse period of adjacent 
35\,d cycles which range from $-6.0\times 10^{-13}$\,s\,s$^{-1}$ to 
$+14\times 10^{-13}$\,s\,s$^{-1}$.
\item Generally, the locally measured $\dot P$ values and the "cycle-to-cycle"
$\dot P$ values show a similar behavior (with the former having the larger
amplitude). For many points, however, the locally measured $\dot P$  
is substantially different from the one derived from $P$ values of 
adjacent measurements (that is from the slope of the pulse period
development).
\end{itemize}
From the last two statements one can conclude that strong pulse period 
variations in \hbox{Her X-1} occur on shorter time scales than the 35\,d 
super-orbital period of the system.

We will discuss the pulse period variations again in Section\,\ref{discussion}.


\section{Correlation between spin-up rate and X-ray flux\label{correlation}}

A positive correlation between the spin-up rate of a neutron star and 
its X-ray luminosity is generally expected from accretion theory 
\citep[see e.g.][]{PringleRees72}, where an increase of the mass accretion 
rate $\dot M$ leads to an increase in the rate of change of the neutron
star's angular momentum. Such a correlation has been observed in many 
accreting pulsars \citep[see e.g.][for a review]{Bildsten_etal97}.
In Her~X-1, however, the correlation was so far questionable \citep[see 
however][]{Wilson_etal94, Klochkov07}. One of the difficulties is the relatively
low amplitude of the $\dot M$ variation in the system (reflected by the 
maximum main-on flux, see below), of a factor $\sim$2. Another problem
is the lack of local spin-up measurements (so far $\dot P$ in 
\hbox{Her X-1} was locally measured only in a few 
cases). For the rest of the existing data $\dot P$ was estimated by taking
differences of the pulse periods between adjacent main-on states which
gives an averaged value over two or more 35\,d cycles 
\citep[see e.g.][where data from \textsl{CGRO/BATSE} were used]{Klochkov07}. 
The X-ray luminosity $L_{\rm X}$ which we assume to be 
proportional to the mass accretion
rate $\dot M$ is usually estimated from the maximum X-ray flux 
during the main-on coincident with the observation
\citep{Staubert_etal07}. Such an estimate is supposed to be very close
to the local value of $L_{\rm X}$. 

\begin{figure}
  \resizebox{\hsize}{!}{\includegraphics{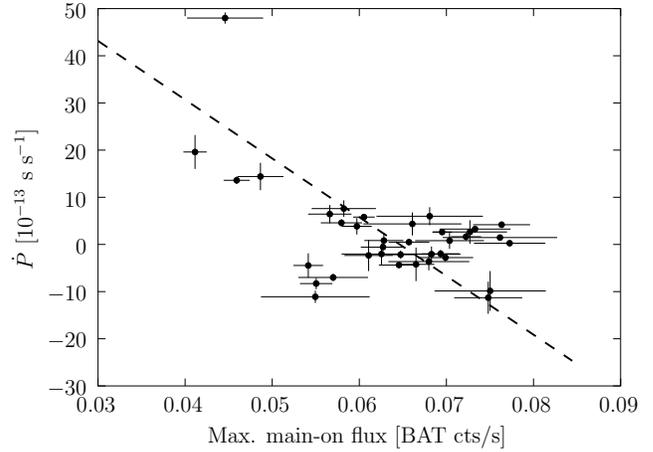}}
  \caption{The locally measured time derivative of the pulse period, $\dot P$,
    in \hbox{Her X-1} versus the maximum X-ray flux during the main-on state
    as determined from the \textsl{Swift/BAT} data. A linear fit (taking
    uncertainties of both variables into account) is shown with the dashed 
    line.}
  \label{cor}
\end{figure}

In this work we have used the values of $\dot P$ locally measured with
\textsl{BAT} to explore the correlation between the spin-up rate
and the X-ray luminosity. Following \citet{Staubert_etal07} we have used
the maximum main-on flux detected with \textsl{BAT}\footnote{We used
the \textsl{Swift/BAT} transient monitor results provided by the 
\textsl{Swift/BAT} team \\
(http://heasarc.gsfc.nasa.gov/docs/swift/results/transients/).} 
as a measure of $L_{\rm X}$. In Fig.\,\ref{cor}
we plot the measured pulse period derivative versus the maximum main-on flux. 
The dashed line indicates a linear fit to the data that takes 
uncertainties of both variables into account (using the orthogonal
regression method, \citealt{Boggs_etal89}). The data indicate an 
anticorrelation as predicted by the basic accretion theory. Inspection
of the linear Pearson's correlation coefficient gives the probability
of $\sim$4$\times 10^{-4}$ to find the measured correlation by chance.
We note, however, that such a high significance appears
mainly due to the group of four points with high spin-down rate and low  
flux (in the
upper left part of the graph in Fig.\,\ref{cor}). The rest of the points
form an uncorrelated "cloud" around $\dot P = 0$. On the other hand, the data
contain no points with large spin-down ($\dot P>10^{12} s/s$) and large 
luminosity (in the upper right part of the graph) that could destroy the
correlation. 

The best-fit slope of the $\dot P(L_{\rm X})$ dependence assuming
a linear relation if found to be 
$\Delta\dot P_{-13}/\Delta F_{\rm BAT} = -(1.2\pm 0.2)\times 10^{3}$,
where $\dot P_{-13} = \dot P/(10^{-13}\,{\rm s/s})$ and $F_{\rm BAT}$ is the 
count rate measured with \textsl{BAT}. 
Assuming a distance of 7\,kpc (see Introduction) one can find that
0.05\,\textsl{BAT} cts\,s$^{-1}$ approximately corresponds to 
2$\times$10$^{37}$\,erg\,s$^{-1}$. Thus, we can rewrite the found
value of the slope as 
$\Delta\dot P_{-13}/\Delta L_{37} = -30\pm 5$\,
where $L_{\rm 37}=L_{\rm X}/(10^{37})$\,erg\,s$^{-1}$.
At a first approximation the value of the slope can be calculated
using the equation:
\begin{equation}
\frac{d I\omega}{dt}=\dot M_x \sqrt{GMR_d} - \kappa_0 \frac{\mu^2}{R_c^3} +
\kappa_1\mu^2 \left(\frac{1}{R_d^3}-\frac{1}{R_c^3}\right)\,,
\label{omegadot}
\end{equation}
where the first term stands for the spin-up from the inner disk radius $R_d$,
the second term describes spin-down due to magnetic coupling beyond the 
corotation radius $R_c$, and the last term takes into account angular 
momentum exchange due to magnetic coupling in the region between 
$R_c$ and $R_d$, $\kappa_0$ and $\kappa_1$ are
numerical coefficients. The inner disk radius is about the Alfv\'en 
radius of the magnetosphere $R_{\rm A}$, which is
determined in the standard way \citep[e.g.][]{PringleRees72}:
\begin{equation}
R_{\rm A}=\left(\frac{\mu^2}{2\dot M\sqrt{2GM}}\right)^{2/7}.
\label{R_a}
\end{equation}
At the corotational regime when $R_d\sim R_c$, the last term in 
Eq.\,(\ref{omegadot}) can be rewritten in the form 
$\kappa_1\mu^23(R_c-R_d)/R_d^4$ and is much smaller than
the first two terms separately, but can be comparable to their
difference at the point of equilibrium, where $R_d\simeq R_c$.
Indeed, from Eq.\,(\ref{omegadot}) one can find that the spin-up and spin-down
torques separately would give $\dot P\sim +$ or $-$$10^{-11}$\,s\,s$^{-1}$, 
whereas on average we observe $<\dot P> \sim 10^{-13}$\,s\,s$^{-1}$.
The Equation.\,(\ref{omegadot}) can be rewritten as 
\begin{equation}
\dot P = -\frac{P^2}{2\pi I}\left(\dot M\sqrt{GMR_c} -
                         \kappa_0\frac{\mu^2}{R^3_c}\right)\,.
\label{pdot:rc}
\end{equation}
and used to calculate the theoretical slope of the $\dot P(L_{\rm X})$
dependence. Here we did not include the third term in Eq.\,(\ref{omegadot}), 
which is by $(R_c-R_d)/R_d$ times smaller than
the spin-up term, and can be ignored when calculating the expected slope of
$\dot P-L_{\rm X}$ relation.
Assuming a radiative efficiency of accretion of $\sim$10\% 
($L_{X}=0.1\dot Mc^2$) we can obtain
$\partial\dot P_{-13}/\partial L_{37} \simeq -50$ which is
close to the value of $\Delta\dot P_{-13}/\Delta L_{37}$
found from the fit of the \textsl{BAT} data ($-30\pm 5$).
A detailed discussion of the observed correlation will be given in 
Sects.\,\ref{corr:discussion} and \ref{eject:discussion}.

\section{Discussion\label{discussion}}

\subsection{Pulse period variations. Comparison with \textsl{BATSE} data\label{per:discussion}}

\begin{figure*}
  \sidecaption
  \includegraphics[width=12cm]{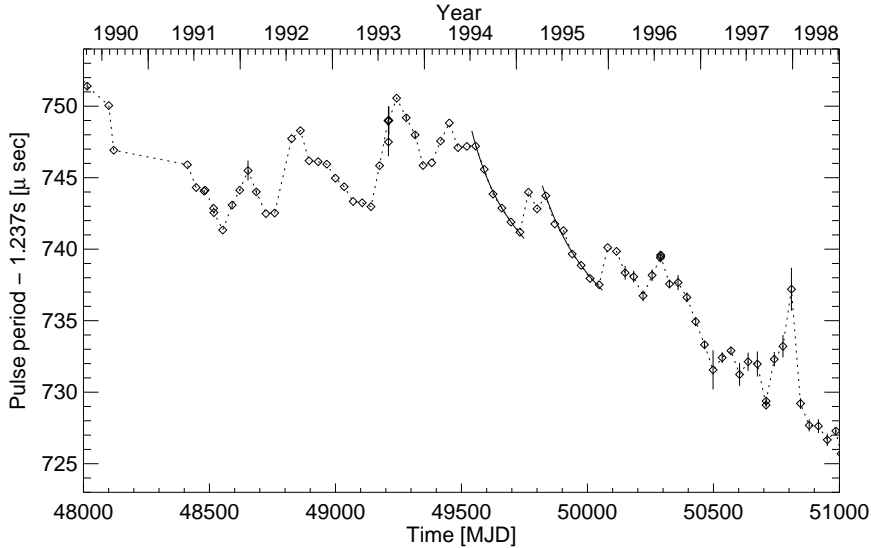}
  \caption{Pulse period $P$ of Her X-1 measured with \textsl{CGRO/BATSE} as 
  a function of time. The values are derived from the publicly available
  pulse search data (R. Wilson, priv. comm.). The solid lines indicate
  the exponential fit to the pulse period during the "relaxation" time after 
  two "glitch"-like episodes.}
  \label{batse}
\end{figure*}

The general behaviour of the pulse period in Her X-1 measured
with \textsl{Swift/BAT} (Fig.\,\ref{per}) is similar to that observed 
previously in this source. The long-term spin-up trend is occasionally
interrupted by short (a few 35\,d cycles) spin-down episodes. 
In Fig.\,\ref{batse} we plot for comparison the historical development of the
\hbox{Her X-1} pulse period measured with \textsl{CGRO/BATSE}.
We have made use of the publicly distributed pulsar data as well as 
lists kindly provided by R. Wilson \citep[see also][]{Staubert_etal06}.

In both, the \textsl{BATSE} and \textsl{BAT} data, the spin-down episodes
are generally shorter than the spin-up episodes. One also sees that in 
most cases the averaged absolute value of $\dot P$ is larger during
the spin-down intervals than during spin-ups. Such an asymmetry is
difficult to explain within the basic accretion theory (i.e. using 
Eq.\,\ref{omegadot}) assuming that the spin period reflects
stochastic variations of $\dot M$. In some cases spin-down
episodes with subsequent spin-ups are reminiscent of 
glitch-like behaviour
observed in some radio-pulsars where a rapid change of the pulse period
is followed by a slower "relaxation" to the long-term trend. 
However, compared to radio pulsars, the "glitches" in \hbox{Her X-1}
appear with the opposite sign: an initial rapid increase of the period
is followed by a slower decrease (as it should be if in both
radio pulsars and \hbox{Her X-1} the glitches are caused by a sudden
decrease/increase of the moment of inertia of the star's crust). 
Two such glitch-like episodes
are indicated in Fig.\,\ref{batse} by solid lines which show the fit
of the "relaxation" intervals by an exponential function on top
of the linear decrease of the period \citep[see e.g.][]{ShemarLyne96}. 
The characteristic relaxation times, $\tau$, for the two cases
are around 100\,days which is similar to the values observed
in radio pulsars \citep{ShemarLyne96}. Some of the other
spin-up/spin-down transitions also resemble glitches, 
even though their decay is less clearly exponential.
One should note, however,
than in radio pulsars the glitch-like variations of
the spin period are predominantly observed in cases of young neutron
stars, with characteristic ages ($\tau = P/2\dot P$) less than 1\,Myr.
The neutron star in \hbox{Her X-1} is believed to be much older.

\subsection{Correlation $\dot P - L_{\rm X}$\label{corr:discussion}}

As we have shown, the \textsl{BAT} data
indicate a correlation between the spin-up rate ($-\dot P$) and
the X-ray flux $L_{\rm X}$ with the slope of the correlation
consistent with the prediction of the basic accretion theory.
However, even for the relatively narrow range of $L_{\rm X}$
observed in \hbox{Her X-1} the correlation appears to be surprisingly
loose. There are several {\em physical} and {\em observational} 
factors that might introduce dispersion in the observed 
$\dot P - L_{\rm X}$ relation. The observational factors include the 
systematics that affects our measurements of the spin-up rate and
the X-ray flux. As it was already mentioned in Sect.\,\ref{correlation},
the pattern-matching technique requires a stable shape of the 
pulse profiles. Even though we tried to restrict our analysis with the
intervals where the pulse profile does not change significantly, we cannot
exclude that the variability of the profile shape
contributes to the measured value of $\dot P$. 
We believe, however, that this systematic effect, if present, 
does not exceed the statistical errors.
On the other hand,
the X-ray luminosity $L_{\rm X}$ might not be unambiguously related
to the maximum main-on flux if the latter is subject to variable 
absorption by a hot corona even in the middle of a main-on
where the angle between the plane of the accretion disk and observer's
line of sight reaches a maximum.

A physical reason for the scattering in the correlation is a
possible bimodal behaviour of the pulsar where it switches between
the spin-up and spin-down regimes at the same level of X-ray luminosity. 
Such possibility appears e. g. in the accretion torque theory
presented by \citet{Lovelace_etal95} where spin-up/spin-down transitions
occur if the corotation radius is close to the Alfv\'en radius
(or, more precisely, to the turnover radius $r_{\rm to}$ introduced by
\citealt{Lovelace_etal95}), i. e.
if the pulsar is close to the so-called corotational regime when 
the spin-up and spin-down torques are nearly balanced. 

The "uncorrelated" cloud
of data points around $\dot P=0$ in Fig.\,\ref{cor} might indicate
the area of the spin-up/spin-down transitions (bimodal behaviour) 
where different values
of $\dot P$ are observed at the same level of $L_{\rm X}$.

Another factor that might increase the dispersion of data points
in the correlation relates to the idea of a freely
precessing neutron star
in \hbox{Her X-1} \citep{Truemper_etal78}. Having a number of
difficulties from the theoretical point of view 
\citep{Shaham77,Sedrakian_etal99}, this hypothesis, however, can explain the 
systematic variation of X-ray pulse profiles exhibited by the system 
\citep[][and references therein]{Staubert_etal09b}. 
Free precession results in 
systematic variations of the observed spin period of the neutron
star \citep{Shakura88,Postnov_etal91,Bisnovatyi_etal93}.  
With the geometrical parameters of free precession that can be assumed 
for \hbox{Her X-1} on the basis of modelling of its pulse profiles 
(K.\,Postnov, priv. comm., paper is in preparation) 
the amplitude of $\dot P$ variation
due to free precession might be as high as $10^{-12}$\,s/s which is 
comparable to the variations that we observe with \textsl{BAT} 
(Fig.\,\ref{cor}).
Since the precessional phase might be different in different
observations, the described effect might introduce additional
scattering in the $\dot P - L_{\rm X}$ correlation if the
free precession indeed takes place in \hbox{Her X-1}. 

\subsection{Evidence for the coronal mass ejection\label{eject:discussion}}

As mentioned in Sect.\,\ref{correlation},
in Her X-1 spin-up and spin-down torques are very well balanced, so
that the averaged $\dot P$ is relatively close to zero.
However, from the points in the upper left part of the graph in 
Fig.\,\ref{cor} one can conclude that there occur extremely 
large spin-down torques (at small fluxes), which are up to 5 times as strong
as spin-up ones. So, the key feature to explain is why do we observe 
such strong spin-down episodes?

One possibility is to assume that the accretion disk carries some
magnetic field which can interact via reconnection with the neutron 
star's magnetosphere beyond the corotation radius $R_c$. This might imply 
that beyond  $R_c$ the field lines can sometimes inflate to become open 
\citep[see also discussion in][]{Lovelace_etal95}.
During such episodes, a substantial fraction of matter in the inner part of
the accretion disk can escape the system in the form of a coronal 
wind ejection along the open field lines. 

Such an ejection of matter should be reflected 
in a secular change of the system's
orbital period which is indeed observed in \hbox{Her X-1} 
\citep{Deeter_etal91,Staubert_etal09}.
To assess the importance of coronal mass ejections for the orbital period 
evolution, we invoke 
general considerations of the non-conservative 
treatment of binary orbital parameters \citep[see e.g.][]{Grishchuk_etal01}.
We shall assume a circular binary orbit. Let $M_o$ be the mass of 
the optical star and $M_x$ -- that of the neutron star, $q=M_x/M_o$ is the 
binary mass ratio. The total angular momentum of the binary is 
mostly stored in the orbital motion of the binary components: 
$J=J_{\rm orb}=M_xM_o/(M_x+M_o)\omega_{\rm orb} a^2$, 
where $\omega_{\rm orb}=2\pi/P_{\rm orb}$ is the orbital angular frequency, 
$P_{\rm orb}$ is the orbital period, and $a=a_x+a_o$ is the binary's 
semimajor axis. Let us define the non-conservativeness parameter in the 
standard way \citep[e.g.][]{RitterKolb92}:
\begin{equation}
\eta=-\frac{\dot M_x}{\dot M_o}\le 1, \quad \dot M_o<0\,.
\end{equation}
Next we assume that the ejected mass carries away the specific angular 
momentum of the neutron star 
$j_x=\omega_{\rm orb} a_x^2=\omega_{\rm orb} a^2 (M_o/(M_x+M_o))^2$. 
(This assumption is justified by the small inner radius of the disk 
$R_d\ll a_x$). Then using the Kepler's 3-rd law and the total angular 
momentum balance, $\dot J = j_x(\dot M_x +\dot M_o)$, 
we can express the fractional change of the orbital period through 
$\dot M_x/M_x$, $q$ and $\eta$:
\begin{equation}
\frac{1}{3}\frac{\dot P_{\rm orb}}{P_{\rm orb}}=-\frac{\dot M_x}{M_x}\left[
1-\frac{q}{\eta}-\left(1-\frac{1}{\eta}\right)\frac{q/3+1}{q+1}
\right]\,.
\label{dPP}
\end{equation}
For $\eta=1$ the familiar expression for the conservative mass exchange 
is obtained. It is convenient to normalize the mass accretion rate onto 
the neutron star $\dot M_x$ 
to the value that can be derived from the observed fractional change of the 
orbital period in the conservative case:
\begin{equation}
\frac{(\dot M_x)_{\rm cons}}{M_x}=
\frac{1}{3}\frac{\dot P_{\rm orb}}{P_{\rm orb}}\frac{1}{q-1}\,,
\label{dotMc}
\end{equation}
so that $\dot m\equiv \dot M_x/(\dot M_x)_{\rm cons}$. 
In \hbox{Her X-1} $q\simeq 0.63$ and 
one finds $(\dot M_x)_{\rm cons}\simeq 8 \times 10^{17}$\,g\,s$^{-1}$ 
for the measured $\dot P_{\rm orb} = -4.85\times 10^{-11}$\,s\,s$^{-1}$
\citep{Staubert_etal09}.
Then we can eliminate $\dot P_{\rm orb}/P_{\rm orb}$ from the left hand side of 
Eq.\,(\ref{dPP}) to obtain the equation for $\eta$ at a given $\dot m$:
\begin{equation}
\eta=\frac{q^2+\frac{2}{3}q-1}{\frac{q^2-1}{\dot m}+\frac{2}{3}q}\,.
\label{eta}
\end{equation}
From here we see that $\dot m<1$ leads to $\eta<1$, i. e. if one wants
to decrease the mass accretion rate onto the neutron star to get 
smaller X-ray luminosity (as is the case of \hbox{Her X-1}, where the mean 
observed $L_x\sim 2\times 10^{37}$\,erg\,s$^{-1}$ is 4 times
smaller than the one following from the conservative mass 
exchange analysis, \citealt{Staubert_etal09}), 
a non-conservative mass exchange is required. Specifically,
if we want to bring in accordance the observed $\dot P_{\rm orb}$ 
and $L_{\rm X}$ in 
\hbox{Her X-1}, we would need $\dot m\simeq 1/4$ and (from Eq.\,\ref{eta}) 
$\eta\sim 0.1$, a fairly high non-conservative mass exchange.

Here we should note that the binary mass ratio $q$ in \hbox{Her X-1} may be 
uncertain. Indeed, recent analysis  of non-LTE effects in the formation 
of the $H_\gamma$ absorption line allows two solutions: $q=0.45$ and $q=0.72$ 
\citep{Abubekerov_etal08}. Eq.\,(\ref{dotMc}) implies that
the critical $(\dot M_x)_{\rm cons}$ decreases for smaller $q$: for $q=0.45$ we obtain
$(\dot M_x)_{\rm cons}\simeq 5.4 \times 10^{17}$\,g\,s$^{-1}$. For this mass ratio 
the appropriate value of the dimensionless parameter $\dot m$ is $1/2$, 
for which from Eq.\,(\ref{eta}) we find $\eta\simeq 0.38$.
This value implies that on average about half of the matter transferred
through the disk should escape from the system to provide the 
observed decrease rate of the system's orbital period.
Note, however, that such a small mass ratio suggests an unusually small 
mass of the neutron star about 0.85 $M_\odot$.

It is very likely that an accretion disk corona is present in \hbox{Her X-1} 
(see recent analysis of \textit{Chandra} X-ray observations 
\citealt{Ji_etal09}). This suggests that there might be a permanent 
coronal accretion disk wind which carries out some angular momentum
from the system.
Episodic ejection of matter in Her X-1 has sporadically appeared in the 
literature to explain some properties of the system. 
\citet{CrosaBoynton80} found that the averaged mass transfer
rate to the outer rim of the accretion disk is somewhat larger
than that required to maintain the observed X-ray luminosity.
In the model of \citet{SchandlMeyer94} the disk wind
results from irradiation of the disk by the central source. 
\citet{Vrtilek_etal01} and \citet{Boroson_etal01} found signatures
of outflowing gas in the UV spectrum of the system.
In the frame of our model, the mass ejection from the system
through the open magnetic field lines occurs most efficiently
during strong spin-down episodes which are associated with small X-ray 
luminosity.  
Indeed, as we see in Fig\,\ref{cor}, the observed 
X-ray flux is decreased by a factor of two during strong spin-down. 
From Eq.\,(\ref{eta}) it is easy to find that
at a given $q$ a fractional decrease in $\dot m$ leads to comparable 
fractional decrease in $\eta$, i.e. accretion indeed becomes more 
non-conservative during strong mass ejection episodes.
During such episodes, the neutron star spin-down power
$I\omega \dot \omega$ is spent to expel accreting matter from the inner disk
radius $R_d\sim R_c$:
\begin{equation}
I\omega \dot \omega \sim \dot M_{\rm ej}\frac{GM}{R_c}\,.
\end{equation}
This equation is satisfied for the observed parameters of \hbox{Her X-1}: 
ejected mass rate during strong spin-downs 
$\dot M_{\rm ej}\sim 0.5\dot M_x\simeq 10^{17}$\,g\,s$^{-1}$, 
$\dot P \simeq 10^{-12}$\,s\,s$^{-1}$, and $R_c\simeq 1.3\times 10^8$\,cm.


\section{Summary}

We have used the publicly available \textsl{Swift/BAT} data to study
the long-term behaviour of the pulsar's spin period in \hbox{Her X-1}.
The measured pulse period variations were compared with those
observed previously with \textsl{CGRO/BATSE}. 

For the first time the pulse period derivative was measured
for a long series of observed main-on states of the source. This allowed
us for the first time to test the correlation between the X-ray 
luminosity and the locally measured spin-up rate of the 
neutron star in \hbox{Her X-1}. We argue that the data indeed show such a 
correlation with the slope consistent with the prediction of the 
basic accretion theory for the parameters of \hbox{Her X-1}.
The relatively large scattering of the data point in the
vicinity of $\dot P\sim 0$ can be caused by the bimodal behaviour of 
the accretion flow configuration at the magnetospheric
boundary which results in switching of the pulsar between spin-up 
and spin-down branches when it stays close to the corotational
regime. In addition, free 
precession of the neutron star, if it takes place in the system, can affect 
significantly the measured values of $\dot P$ and, therefore, contribute
to the scattering.

We argue that together with the long-term decrease of the 
orbital period in \hbox{Her X-1} the measured pulse period behaviour 
requires the presence of mass ejection from the inner parts of the
accretion disk along the open magnetic field lines. The mass ejection
episodes probably take place during strong spin-down episodes which are 
associated with the small X-ray luminosity.

We would like to note that the described technique allows one
to use the \textsl{BAT} instrument as a long-term monitor
of spin periods in other bright accreting pulsars
\citep[see also][]{Camero_etal09}. 

\begin{acknowledgements}
The work was supported by the DLR grant BA5027, RFBR grant 09-02-00032,
and DFG grants Sta 173/31 and RUS 113/717/0-1

We thank ISSI (Bern, Switzerland) for its hospitality during the 
team meetings of our collaboration when we discussed the presented
results.

DK thanks Valery Suleimanov (IAAT, T\"ubingen) for useful discussions.
\end{acknowledgements}

\bibliographystyle{aa}
\bibliography{ref}

\end{document}